# Role of the electron-phonon coupling in tuning the thermal boundary conductance at metal-dielectric interfaces by inserting ultrathin metal interlayers

Shany Mary Oommen[1], Simone Pisana[1,2,*]

[1] Department of Physics and Astronomy, York University, Toronto, Canada

[2] Department of Electrical Engineering and Computer Science, York University, Toronto, Canada

[*] Corresponding email: pisana@yorku.ca

## Abstract

Varying the thermal boundary conductance at metal-dielectric interfaces is of great importance for highly integrated electronic structures such as electronic, thermoelectric and plasmonic devices where heat dissipation is dominated by interfacial effects. In this paper we study the modification of the thermal boundary conductance at metal-dielectric interfaces by inserting metal interlayers of varying thickness below 10 nm. We show that the insertion of a tantalum interlayer at the Al/Si and Al/sapphire interfaces strongly hinders the phonon transmission across these boundaries, with a sharp transition and plateau within ~1 nm. We show that the electron-phonon coupling has a major influence on the sharpness of the transition as the interlayer thickness is varied, and if the coupling is strong, the variation in thermal boundary conductance typically saturates within 2 nm. In contrast, the addition of a nickel interlayer at the Al/Si and the Al/sapphire interfaces produces a local minimum as the interlayer thickness increases, due to the similar phonon dispersion in Ni and Al. The weaker electron-phonon coupling in Ni causes the boundary conductance to saturate more slowly. Thermal property measurements were performed using time domain thermo-reflectance and are in good agreement with a formulation of the diffuse mismatch model based on real phonon dispersions that accounts for inelastic phonon scattering and phonon confinement within the interlayer. The analysis of the different assumptions included in the model reveals when inelastic processes should be considered. A hybrid model that introduces inelastic scattering only when the materials are more acoustically matched is found to better predict the thickness dependence of the thermal boundary conductance without any fitting parameters.

## Introduction

Interfaces play an important role in nanoscale heat transport. In thin films with thickness of the order of the Kapitza length, interfaces cannot be neglected. The Kapitza length can be expressed as $l_K= R_K k$, where the Kapitza resistance $R_K=1/G$, is the inverse of the thermal boundary conductance $G$ at the film's interface, and $k$ is the thermal conductivity of the film. The thermal boundary conductance relates the temperature drop $\Delta T$ at interfaces to the heat flux $J$ crossing it, $G=J\Delta T$. Highly integrated electronic structures are often composed of numerous thin films with dissimilar properties, and therefore detailed knowledge of heat transport at these interfaces is required in numerous significant technologies [1–4]. For example, highly efficient thermoelectric devices require a low $k=k_e+k_p$, where the total thermal conductivity has components due to phonons ($k_p$) and electrons ($k_e$) [5–7]. Since $\sigma$ and $k_e$ are interrelated, reducing the phonon thermal conductivity of the thermoelectric material is the most preferred way for enhancing the thermoelectric efficiency [8,9]. Another example is the heat generation in heat-assisted magnetic recording (HAMR) devices [10]. Here, a plasmonic near field transducer (NFT) heats a nearby magnetic medium by concentrating laser energy to a sub-wavelength spot. To avoid excessive heating of the plasmonic device, the heat generated in the NFT metal must be efficiently dissipated to a nearby dielectric through a boundary. This has to be achieved while preserving low optical losses in the plasmonic device, but typical low-loss plasmonic metals have a high thermal boundary resistance, so the introduction of a higher-loss material to improve the thermal performance has to be kept to a minimum.

Understanding interfacial heat transport mechanisms can provide additional ways to tune the thermal boundary conductance, which is critical in heat management applications and for the engineering of novel micro and nano-electronic devices [6,9–11]. Cheaito *et al.* studied both experimentally and theoretically the thermal boundary conductance accumulation function across a range of metal/native oxide/Si and metal/sapphire systems, showing the importance of phonon spectral overlap to obtain large-$G$ interfaces [12]. Despite the existing body of literature reporting on the thermal boundary conductance $G$ of various interfaces, tuning or modifying it remains a major challenge. The addition of a metallic interlayer is an area of active research [13–17]. The dominant factors that drive the change in $G$ with the use of a metallic interlayer are believed to be: intermediate phonon properties to bridge mismatch between top metal and dielectric, influence of the electron-phonon coupling constant $g$ of the interlayer, and changes in metal-dielectric bond strength. This highlights differing contributions to $G$, namely the phonon properties of the materials adjacent the interface, the rate of energy transfer between electron and phonons, and chemical/structural properties of the interface. The latter factor is the least investigated, given the difficulty in experimentally quantifying adhesion strength as function of metal and interlayer composition.

Wang *et al.* [13] used Boltzmann transport simulations to suggest that the insertion of an interlayer with intermediate $g$ strength at a metal-dielectric interface significantly enhances $G$. They theoretically studied the effect of Al and Pt interlayer at the Au/Si interface with interlayer thickness varying between 10 nm to 100 nm.

Having a $g$ in the interlayer larger than that of the top Au metal allows for a more conductive energy transfer pathway from the electrons, which are the dominant energy carriers in the top metal and the interlayer, to the phonons, which carry the energy in the dielectric substrate. Similarly, Li *et al.* [14] showed that insertion of a 20 nm thick Ni interlayer at Au/Sapphire interface can reduce the interfacial resistance by 70%, though its thickness dependence was not studied. Jeong *et al.* [15] showed the effect of adding an interlayer with intermediate Debye temperature between a metal and dielectric by studying the thickness-dependent effect of Cu and Cr interlayers at the Au/sapphire interface. Adding a material with intermediate Debye temperature can enhance thermal boundary conductance by bridging phonon transport, since otherwise the elastic phonon scattering phase space would be reduced. This reasoning is in line with non-equilibrium molecular dynamics simulations of English *et al.* [16]. Jeong *et al.* [15] also developed a model to predict the evolution of $G$ as a function of interlayer thickness based on phonon-metal/phonon-dielectric transport. Their model assumed that the phonons from Au pass directly to the substrate when the interlayer thickness is smaller than phonon wavelength, whereas phonons with wavelength smaller than interlayer thickness come directly from the interlayer. They ignored the effect of the electron-phonon coupling in both metal and interlayer. Blank and Weber [17] studied the thickness dependence of a Cu interlayer by developing a model that accounts for both phonon-phonon interactions and electron-phonon coupling in the interlayer. With the experimental values for $G$ as a starting point in the limit of zero interlayer thickness and fitting for $g$ of the interlayer, their model reproduced the evolution of $G$ with thickness in Au/sapphire, Au/diamond, and Au/Si systems.

In this work, we study the effect of a Ni or Ta interlayer at the Al/sapphire and Al/Si interfaces. By choosing vibrationally matched or mismatched interlayers with differing values of $g$, we aim to further elucidate the relative importance of these factors in how $G$ evolves for interlayer thicknesses below 10 nm. We are concerned with interlayer thickness less than 10 nm because for applications such as plasmonic devices, it is important to tailor heat dissipation without introducing materials that give rise to optical losses, so the interlayer thickness needs to be as small as possible. Few studies in the literature focus on how $G$ is affected by interlayer thicknesses below 10 nm [15,17]. Ni was chosen as an interlayer because it has relatively similar acoustic properties to that of Al and the substrates, but has higher $g$ than Al. In the case of Ta, it has dissimilar acoustic properties with respect to Al and the substrates, but Ta has a value for $g$ even higher than both Ni and Al. We will show that the value of $g$ has a major influence on the sharpness of the evolution of $G$ with interlayer thickness, and we will test the influence of the various assumptions used to model this behavior. A new hybrid approach to the diffuse mismatch model will be introduced to better describe the thickness dependence of the interlayer material even with interfaces with mismatched phonon properties.

**Experiment**

We deposited Al as the top metal and interlayers (Ni or Ta) on c-sapphire (0001) and Si substrates by dc magnetron sputtering in an argon atmosphere with a base pressure of $\sim 1\times10^{-7}$ Torr. The depositions were carried out at 200 W and 3 mTorr with rates for Al, Ni and Ta of 1.1 Å$s^{-1}$, 1.4 Å$s^{-1}$ and 1.8 Å$s^{-1}$, respectively. Prior to the metal bilayer deposition, the substrates were cleaned by sonicating in acetone and isopropanol (~10 min each) using an ultrasonic bath. The thickness of the metal bilayers was determined by the deposition rates calibrated through picosecond acoustics and profilometry of thicker reference samples and checked for every deposition run. The Al thickness was measured to be ~50 nm for all samples, and the interlayer thickness was varied between 0.25 nm to 10 nm. All resulting film thickness values have an uncertainty of 10 %.

The thermal properties of the metal-interlayer-substrate system were measured using a time domain thermoreflectance (TDTR) set-up [18,19]. This technique uses an ultrafast pump beam to heat up the sample surface, and an ultrafast probe beam probes the change in reflectivity as a function of time, thereby sampling the changes in surface temperature through thermoreflectance. The relative arrival time of the pump and probe beams is controlled by an optical delay stage, either in the pump or probe beam path. We use an ultrafast fiber laser (Amplitude Laser, Satsuma HP$^2$) that generates pulses centered at 1030 nm at a repetition rate of 40 MHz. We use two wavelengths for the probe and pump beams where the latter has been frequency doubled from 1030 nm to 515 nm. The pump beam passes through an electro-optic modulator (Conoptics, 350-160) to impose a square wave modulation at 1 MHz to facilitate lock-in detection (Zurich Instruments, HF2LI). Pump and probe beam spot sizes were measured to have $1/e^2$ radii of 7.5 μm and 4.1 μm, respectively, using a knife-edge technique. The thermal properties of interest are extracted by fitting the experimental data (thermal phase as function of pump-probe time delay) to the solution of the heat diffusion equation for a periodic point source on a semi-infinite layered media [20]. We fit for the thermal boundary conductance $G$ of the metallic-bilayer/substrate interface and the thermal conductivity $k_{sub}$ of the substrate. The extracted $k_{sub}$ matches the bulk literature values within the experimental uncertainty (132 W/mK for Si and 32 W/mK for sapphire) and varies by less than 1 % for all samples. For fitting purposes, the metallic bilayer is considered as a single layer having a thickness equal to the total bilayer thickness and volumetric heat capacity equal to the thickness-weighted value. The thermal conductivity of the bilayer was obtained through the Wiedemann-Franz Law and 4-point-probe measurements of the electrical conductivity of the metallic films. All volumetric heat capacity values were obtained form the literature. The TDTR data and best fits for all samples measured in this work is presented in the Supplementary Information. In order to estimate the error propagated in the fits from experimental noise or uncertainties in thermophysical parameters that were held constant, a Monte Carlo approach described previously was used [21]. Briefly, each TDTR scan is fit for *G* 300 times, and at each iteration random normally-distributed variations in some assumed values were used to determine how the fit results vary due to errors in the assumed values. The parameters that were varied in the Monte Carlo iterations were those for which there was larger uncertainty. The parameters and associated uncertainties were: the thickness of the metal layer (10 %), thermal conductivity of the metal layer (10 %), optical spot size (2 %), thermal conductivity of the substrate (8% for Sapphire and 11% for Silicon) and phase noise of the TDTR measurement (0.1 to 0.5 degrees, estimated from the phase noise in each measurement). The resulting variations in *G* for the Monte Carlo iterations yielded the error bars that are smaller than the symbol size in the figures. Additional details pertaining to the uncertainty analysis of the fits and Monte Carlo simulations are provided in the Supplementary Information.

**Modeling of *G* for the metal-interlayer-substrate system**

At metal-dielectric interfaces both phonon and electron transport channels should be considered, together with the electron-phonon coupling *g* which governs the rate at which energy is transferred between these channels [22] and affects the overall thermal boundary conductance *G*. In the presence of an interlayer, several energy transfer pathways exist among the channels. Following the approach of Blank and Weber [17], we consider the total *G* involving two parallel pathways: *pathway 1* and *pathway 2* (Figure 1). The energy transfer begins with the laser-excited electrons in the top metal and ends with phonon transport in the dielectric substrate. We ignore interfacial electron-phonon coupling, as this was shown to play a role only at high electron temperatures, a regime not reached in these experiments [19,23,24]. Indeed, estimates of electron temperature rise in our experimental conditions using a two-temperature model reveal that the electron temperature rise is well below 50 K, which is lower than the lattice temperature (300 K). This is opposite the requirement of electron temperature rise being much greater than the lattice temperature in order to observe an appreciable influence of interfacial electron-phonon coupling [19,23,24].

In *pathway 2* the electrons in the top metal interact with the electrons in the metallic interlayer to transfer energy ($G_{ee}$). After this, the interlayer electrons transfer their energy to the lattice via electron-phonon coupling ($G_{ep,2}$). Finally, the phonons in the interlayer transfer energy to phonons in the substrate ($G_{pp,2}$). Given the nanometric thickness of the interlayer, we only consider phonons with wavelength smaller than the interlayer thickness to participate in this pathway [15,17]. This is because phonons with wavelength larger than the interlayer thickness would conceptually extend beyond the layer's physical size and would have hybrid characteristics determined by the vibrational properties of the interlayer and adjacent materials. This would make the vibrations spectrally varied and the concept of a phonon would be meaningless. Therefore, only phonons with sufficiently small wavelength are considered in this layer, with their properties determined by the composition of the interface material. Note that this assumption does not entirely exclude phonons with wavelength larger than the interlayer size in the overall interface transport, as long wavelength phonons will be considered in *pathway 1* as described below. The addition of an interlayer with an electron-phonon coupling constant higher than that of the top metal layer can result in back transfer of heat from the interlayer to the top metal layer ($G_b$). This occurs because the phonon temperature in the interlayer is larger than that of the top metal. This effect however only lasts for a few picoseconds, as the electron and phonon temperatures equilibrate in both the top metal and interlayer. We analyze the system response starting from 100 ps after the arrival of the pump pulse, thus the contribution of $G_b$ to the overall conductance measured can be ignored [19]. The total interfacial conductance due to *pathway 2* can be expressed assuming the resistance due to all three energy transfer steps are in series as:

$$\frac{1}{G_2} = \frac{1}{G_{ee}} + \frac{1}{G_{ep,2}} + \frac{1}{G_{pp,2}} \tag{1}$$

In *pathway 1*, electrons couple with the phonons in the metal ($G_{ep,1}$), and subsequently the phonons exchange energy with the substrate ($G_{pp,1}$). Phonons from the metal layer pass directly to the substrate if the phonon wavelength is larger than the interlayer thickness, as done previously [15,17]. The total interfacial conductance due to *pathway 1* can again be expressed assuming the resistances are in series as:

$$\frac{1}{G_1} = \frac{1}{G_{ep,1}} + \frac{1}{G_{pp,1}} \tag{2}$$

In the following we will outline how the components of Eqs. (1, 2) can be treated, including considerations for finite size effects, and finally we will combine the results into a total value for $G$ for both pathways.

To model the phonon thermal transport at the interface between metal and dielectric ($G_{pp,1}$ and $G_{pp,2}$) we use a modified diffuse mismatch model (DMM) [25]. We consider realistic phonon dispersion relations to calculate the phonon transmission coefficient and $G$ [26], and assume an isotropic phonon dispersion along the crystal growth direction [15]. Assuming an isotropic phonon dispersion makes the computation less time consuming and is a good enough approximation of the three-dimensional phonon dispersion [25]. We use different DMM scattering models (in the limits of purely elastic scattering or considering all elastic and inelastic scattering processes) and the role of including optical phonons to check the importance of this assumption on the systems studied here [27]. For elastic scattering, the phonon transmission probability from material A to B can be expressed as [25]:

$$\alpha_{A\to B} = \frac{\sum_{j,B} \hbar\omega_{j,B} q_{j,B}^2 \upsilon_{j,B} f_{BE}}{\sum_{j,B} \hbar\omega_{j,B} q_{j,B}^2 \upsilon_{j,B} f_{BE} + \sum_{j,A} \hbar\omega_{j,A} q_{j,A}^2 \upsilon_{j,A} f_{BE}} \tag{3}$$

On the other hand, for the case of all elastic and inelastic scattering processes, the phonon transmission probability from material A to B can be expressed as [25]:

$$\alpha_{A\to B} = \frac{\sum_{j,B} \int \hbar\omega_{j,B} q_{j,B}^2 \upsilon_{j,B} f_{BE} dq_{j,B}}{\sum_{j,B} \hbar\omega_{j,B} q_{j,B}^2 \upsilon_{j,B} f_{BE} dq_{j,B} + \sum_{j,A} \hbar\omega_{j,A} q_{j,A}^2 \upsilon_{j,A} f_{BE} dq_{j,A}} \tag{4}$$

In the above expressions, $j$ refers to the phonon branch, $\hbar$ is the reduced Plank constant, ω is the phonon angular frequency, $q$ is the phonon wave vector, $\upsilon$ is the phonon group velocity, $f_{BE} = 1/\left[\exp\left(\hbar\omega_{j,B}/k_B T\right) - 1\right]$ is the Bose-Einstein distribution, $k_B$ is Boltzmann's constant, and $T$ is the temperature. The phonon dispersion of each branch $\omega_j(q)$ is obtained by using a polynomial fit to the experimental dispersion curves along the preferential growth direction reported for our substrates. We choose $\Gamma \to X$ $(001)$ for Al, $\Gamma \to L$ $(111)$ for Ni and $\Gamma \to N$ $(011)$ for Ta [28,29]. The dispersions for the substrates follow their orientation: $\Gamma \to Z$ $(0001)$ for $c$-sapphire, and $\Gamma \to X$ $(001)$ for Si [30,31]. The summation is performed either over all phonon branches or for the acoustic branches only, to determine the relative importance of optical phonons. We note that although various expressions have been presented in the literature to treat elastic and inelastic scattering, the choice of which to use is not trivial. We will comment on this choice further in the Results and Discussion section below, and propose a hybrid elastic/inelastic treatment of interfaces with interlayers based on the relative phonon mismatch among the materials.

The total thermal boundary conductance can be expressed as a first order derivative of total heat current density with respect to temperature:

$$G_{pp} = \frac{1}{8\pi^2} \sum_{j,A} \int_{k_{j,A}} \hbar\omega_{j,A} q_{j,A}^2 \left|\upsilon_{j,A}\right| \alpha_{A\to B} \frac{df_{BE}}{dT} dq_{j,A} \quad (5)$$

Eq. (5) can be used to calculate the total $G_{pp}$ at interfaces, but this expression does not directly allow to limit the calculation to only a fraction of the phonon spectrum. Based on the discussion above, we would like to consider the fraction of phonons with wavelength $\lambda$ less than or equal to the interlayer thickness $h$, thus we would like to limit the integration to $\lambda_{max} = h$. This is more easily accomplished by changing the integration variable from $q$ to $\lambda$ using the relation $q = 2\pi/\lambda$. The accumulation of thermal boundary conductance [12,15] as a function of phonon wavelength can then be expressed as:

$$G(\lambda < \lambda_{max})_{pp,accum} = -\frac{1}{8\pi^2} \sum_{j,A} \int_{\lambda_{min}}^{\lambda_{max}} \hbar\omega_{j,A} q_{j,A}^2 \left|\upsilon_{j,A}\right| \alpha_{A\to B} \frac{df_{BE}}{dT} \frac{2\pi}{\lambda^2} d\lambda_{j,A} \quad (6)$$

where $\lambda_{min}$ represents the shortest wavelength phonon at the Brillouin zone edge and $\lambda_{max}$ is the limit to phonon wavelength we would like to impose. The thickness dependent $G(h)_{pp,1}$, which for *pathway* 1 should not include phonons with wavelengths that can exist in the interlayer (those are considered in *pathway 2*) can therefore be calculated using [15]:

$$G(h)_{pp,1} = G_{pp,sat} - G(\lambda_{max} = h)_{pp,accum} \quad (7)$$

where $G_{pp,sat}$ is the saturated value of $G_{pp,accum}(\lambda < \lambda_{max})$ as $\lambda_{max} \to \infty$. The shape of the $G(h)_{pp,1}$ curve decreases with increasing interlayer thickness, since the contribution to the total $G$ from phonons originating in the metal and transmitting into the substrate decreases as more of the phonons are cut-off.

In a similar way, for *pathway 2*, where the contribution of phonon conduction from the interlayer into the substrate increases with interlayer thickness as more phonon wavelengths are allowed, the thickness dependence is given by

$$G(h)_{pp,2} = G(\lambda_{max} = h)_{pp,accum} \quad (8)$$

The expressions above account for finite size effects by forbidding interlayer phonon wavelengths longer than the interlayer thickness, but this phenomenological treatment does not account for changes in phonon band structure in thin slabs, and still assumes a bulk-like phonon dispersion. We have modeled by lattice dynamics the changes to the phonon band structure in Silicon ultra-thin slabs due to phonon confinement and have found that the phonon band bending near the Brillouin zone center or the creation of quasi-optical bands in slabs as thin as 0.54 nm alter the value for $G_{pp}$ obtained by the DMM by only 6% in a Silicon/Germanium interface. This allowed us to isolate the effect of nanostructure band structure modification and estimate its importance. While this treatment is not as complete as those reported by molecular dynamics, non-equilibrium Green's function or Boltzmann transport approaches, it provides us some confidence that in our DMM framework the tunneling and accumulative effects in phonon transport outlined above are more dominant than phonon confinement effects. This is supported by the results of conductance as function of slab thickness obtained by the Green's function approach [32].

We now turn to the remaining terms contributing to the heat conduction pathways. The thickness-dependent resistance due to the electron-phonon coupling in an interlayer with thickness $h$ can be evaluated by [33]:

$$G_{ep,2} = hg \tag{9}$$

This expression is valid for sub-nanometer thick layers, where electron-phonon coupling is incomplete. The contribution from electron-phonon coupling in the top metal layer (~50 nm thick) can be calculated using $G_{ep,1} = \sqrt{k_p g}$, where $k_p$ is the phonon contribution to the thermal conductivity of the metal [22]. In our work the contribution of $G_{ep,1}$ is negligible, since Al has a high value for $g$, and there is small electron-phonon non-equilibrium (negligible resistance) in top Al layer. The thermal boundary conductance at the interface between two metals due to electron-electron interaction can be expressed as [34]:

$$G_{ee} = \frac{\gamma_{e,A}\nu_{e,A}\gamma_{e,B}\nu_{e,B}}{4(\gamma_{e,A}\nu_{e,A}+\gamma_{e,B}\nu_{e,B})}T \qquad (10)$$

where $\gamma_{e,A} = C_{e,A}/T$ is the Sommerfeld parameter for metal A, $C_{e,A}$ is the electronic heat capacity and $\nu_{e,A}$ the fermi velocity.

The total interfacial conductance predicted by this framework in the presence of an interlayer can be calculated assuming the resistances due to the two pathways are parallel to each other. The total contribution is therefore:

$$G_{model} = G_1+G_2 \qquad (11)$$

As the interlayer thickness increases and becomes comparable to the largest phonon wavelength in interlayer, the contribution from $G_1$ in *pathway 1* becomes negligible. On the other hand, the contribution from $G_2$ in *pathway 2* becomes larger with interlayer thickness, with the growth rate being dominated by $G_{pp,2}$ and $G_{ep,2}$, since the values for $G_{ee}$ are typically large. In the results that follow, the relative contributions of each pathway are shown, together with the DMM results obtained using different phonon scattering assumptions or the inclusion of optical phonons.

## Results and Discussion

### *Samples without interlayer*

We present first the measured thermal boundary conductance of the reference Al/Si and Al/sapphire systems without the Ni and Ta interlayers, and summarize the results in Table 1. The thermal boundary conductance in the absence of interlayers for Al/Si and Al/Sapphire was measured to be 250 $MWm^{-2}K^{-1}$ and 200 $MWm^{-2}K^{-1}$, respectively. The $G$ value measured ($G_{expt}$) are in good agreement with our previous reports [35,36] and in general are in line with literature values [4,12,37,38], although some variation in the reported values can be attributed to differences in residual impurities on the surface of the substrate. The $G$ values predicted from the model ($G_{model}$) for Al/Si and Al/sapphire are 245 $MWm^{-2}K^{-1}$ and 235 $MWm^{-2}K^{-1}$, respectively when considering all inelastic processes and acoustic phonons only, whereas the values increase to 290 $MWm^{-2}K^{-1}$ and 410 $MWm^{-2}K^{-1}$, respectively when considering also the optical phonon contribution. A comparison between $G_{expt}$ and $G_{model}$ indicates that optical phonons in sapphire and silicon contribute very little to the total thermal boundary conductance, particularly for sapphire. This contrasts with the differences produced by the DMM model when optical phonons are considered, as the $G_{model}$ increases by 18% and 74% for Al/Si and Al/sapphire, respectively. On the other hand, the difference in $G_{model}$ when only elastic scattering is considered is not pronounced, since there is almost no energy overlap between the Al phonon dispersion and the optical branches in Si or sapphire (see Figure 6). The results lead us to conclude that even though the inclusion of optical phonons makes a marked difference in the inelastic DMM result, their actual contribution to heat transport is not as pronounced, and an inelastic treatment of acoustic phonon modes is sufficient to explain the results. The minimal contribution of optical phonons is not caused by a low heat flux contribution due to the relatively flat optical bands, since the model would predict a significant enhancement in $G$ when these are considered [25]. While we don't exclude that optical modes may participate somewhat in the heat transport, particularly where the optical mode energies are not significantly above the acoustic ones, their incorporation within the DMM framework tends to overestimate their contribution. The overestimation likely originates from the way inelastic scattering is modeled, where $n$-phonon processes are considered equally as likely, contrary to expectations. Other implementations of the inelastic DMM have been developed to ensure phonon number conservation (see for example [39]), but these are not treated in this work.

***Nickel interlayer***

Figure 2 shows the evolution of *G* as measured and as predicted by the thermal model, as a function of Ni interlayer thickness for the Al/Ni/sapphire system. As the Ni thickness increases, *G* has a local minimum at around ~0.75 nm. The presence of the local minimum can be attributed to the similar vibrational properties of Ni and Al (Table 2) and the higher electron-phonon coupling strength of Ni ($g = 0.36 \times 10^{18}$ $Wm^{-3}K^{-1}$) with respect to Al ($g = 0.23 \times 10^{18}$ $Wm^{-3}K^{-1}$). Al/sapphire and Ni/sapphire have similar $G_{pp}$ due to the similar vibrational properties of Al and Ni. Hence, at higher and lower interlayer thicknesses the *G* values are similar and approach the bulk value of the Ni/sapphire and Al/sapphire interfaces, respectively. At lower Ni thicknesses, *pathway 1* dominates over *pathway 2* (higher *G*) because of two reasons: 1) low phonon contribution from the interlayer due to its low thickness having few allowable phonon modes makes $G_{pp,2}$ small, and 2) low conductance through the electron-phonon coupling in Ni, again due to the low Ni thickness, makes $G_{ep,2}$ small. These low conductances acts in series reducing the overall conductance of *pathway 2*. As the Ni thickness increases, the phonon conduction contribution of the interlayer increases as does the conductance from the electron-phonon coupling. At the same time the conduction through *pathway 1* decreases, since more phonon modes are excluded from $G_{pp,1}$. Figure 2 shows the evolution of the model as a function of thickness when including (black line) or excluding (green line) the electron-phonon coupling contribution of the interlayer. It is clear that the local minimum originates from the electron-phonon coupling in the interlayer, which increases the resistance in *pathway 2*. This overall evolution of the model is in good agreement with the experiment.

In order to check the influence of the assumptions in the DMM model, we performed the DMM analysis assuming elastic and inelastic scattering at the interfaces. Figure 2 compares the evolution of the modeled *G* under differing assumptions. The black curve considers inelastic interactions with the acoustic phonons in sapphire, whereas the pink line considers elastic interactions only. The blue and red curves instead compare the inelastic or elastic interactions of all phonon modes, including the optical branches in sapphire. As expected, including optical modes has a negligible role in the elastic models, given the low energy overlap in the dispersions (see Figure 6). Therefore, going forward we will not further discuss the difference obtained by including optical modes in the elastic models. As seen in Figure 2 for the acoustic-only models, harmonicity is not a dominant consideration, given the similarity of the Al and Ni phonon dispersions to the acoustic modes in sapphire. Considering inelastic contributions for acoustic-only branches increases somewhat the overall *G*, and the experimental data is in reasonable agreement with either of the acoustic-only models. Accounting for optical phonons in the limit of inelastic interactions greatly overestimates the value of *G*. This will be seen for all the data presented here and is again attributed to the equal contribution given to all *n*-phonon processes in the model.

A similar trend was observed for the Al/Ni/Si system, as shown in Figure 3, where the large interlayer thickness Ni/Si interface limit is similar to the Al/Si interface, with a local minimum. The thickness dependent *G* predicted from the model has a local minimum at ~1.5 nm, whereas the experiment shows a local minimum at ~0.25 nm. The accumulation of the phonon wavelength dependence in the model starts only at Brillouin zone edge, so any transition below its equivalent phonon wavelength (0.112 nm) is not captured. Near the Brillouin zone edge the phonon flux is also low, due to the low phonon group velocities, therefore the minimum observed at ~0.25 nm cannot be explained within this model. Interfacial adhesion effects are not captured in the model, and may play an important role in the rapid evolution of the thickness dependence for the Al/Ni/Si system. The acoustic inelastic model agrees reasonably well with the saturated experimental *G* values for Al/Si and Ni/Si.

Considering acoustic branches only in inelastic models for both the Al/Ni/sapphire and Al/Ni/Si systems leads to considering *n*-phonon processes with energy differences less than a factor of two. As we will see in the next section, where the acoustic inelastic model includes phonon energy differences greater than two, the predicted *G* overestimates the experimental data and the elastic model will be shown to be more accurate.

***Tantalum interlayer***

When a Ta interlayer was inserted between Al/sapphire (Figure 4) and Al/Si (Figure 5), we observed a fast monotonic decrease in thermal boundary conductance with increasing thickness. The fast saturation (within 2 nm) in the Al/Ta system can be attributed to the strong electron-phonon coupling in Ta ($g = 31 \times 10^{18}$ $Wm^{-3}K^{-1}$). The strong *g* in Ta reduces the electron-phonon coupling resistance by readily dragging the electron and phonon baths into thermal equilibrium. Figure 4 shows that relatively stronger *g* in Ta with respect to Ni induces a fast saturation of *G* as a function of interlayer thickness. Thus, we can conclude that as the *g* strength increases, the thermal boundary conductance saturates faster as a function of thickness. This will be further illustrated in the next section. The large differences in the vibrational properties of Ta and Al (Table 2) introduce a mismatch at the interface, hindering phonon transmission, causing the saturated value for *G* to be very small. As can be seen in Figure 6, Ta has the worst phonon branch overlap with Si and sapphire, which in turn reduces the phonon flux at the interface. For both Al/Ta/sapphire and Al/Ta/Si systems, the *G* value reached a plateau at 70 $MWm^{-2}K^{-1}$. For both experiments, the model captures a sharp decrease in thermal boundary conductance with a plateau starting at ~1 nm for Al/Ta/sapphire and ~2 nm for Al/Ta/Si. For the Ta interlayer, optical phonons in sapphire or Si do not contribute to the value of $G$ due to the large energy difference between highest acoustic phonon in Ta and lowest optical phonon in the dielectric. We also note that only the elastic model accurately reproduces the saturated value of $G$. Contrary to the Ni interlayer case where an inelastic acoustic model seemed adequate, the Ta acoustic modes are more than a factor of 2 lower in energy than the lowest optical modes in the dielectric, and an inelastic model would likely overestimate the contributions of $n$-phonon processes with large upconversion frequencies. Figure 5 is particularly interesting in this respect, as it shows the acoustic inelastic model reproducing the Al/Si limit (0 nm Ta) where anharmonicity does not involve large energy differences, whereas when Ta is introduced, the elastic model reproduces the trend, since the inelastic model would involve $n$-phonon processes with large energy differences. In order to combine these two treatments, we calculate $G_{pp}$ for Al/Si using the inelastic model and for Ta/Si using the elastic model. Such a hybrid elastic/inelastic model, which introduces inelastic scattering only when the materials are more acoustically matched, reproduces the data well.

***Role of the electron-phonon coupling constant in the interlayer***

In this section we discuss further the role of $g$ in the interlayer and compare its effect with the accumulation of the phonon transport channel $G_{pp,2}$. We used the above-mentioned formulation to predict the $G$ evolution for the data published by Jeong *et al.* [15], who reported on the rapid change of $G$ with thickness in the Au/Cr/Sapphire system and first proposed the accumulation model.

Figure 7 shows the comparison of the model and data of Jeong with our models. Au and sapphire have greatly mismatched vibrational properties like Ta and sapphire. Inserting a Cr interlayer enhances the $G$ by bridging the phonon energies at the interface. As can be seen in the figure, models that don't include the effect of electron-phonon coupling are inaccurate in reproducing the transition around 1-5 nm. Cr has a comparatively low *g*, thus inducing a slower saturation in *G* with respect to thickness at ~5 nm, as opposed to the faster saturation behaviors seen in Figures 2-5. The plot clearly suggests that electron-phonon coupling in the interlayer should not be neglected, and when weak it causes a slower saturation in the thickness dependence of the thermal boundary conductance. Similarly to the case of Al/Ta/Si above, we use a hybrid elastic/inelastic model for the Au/Cr/Sapphire system, where the Au/Sapphire transport is treated elastically due to the high mismatch in phonon energies and the Cr/Sapphire transport is treated inelastically. Although our model reproduces the data and slow saturation relatively well, it appears that the value of *g* for Cr of $0.42 \times 10^{18}$ $Wm^{-3}K^{-1}$ is smaller than what the thickness trend would suggest, and an increased value of $10^{18}$ $Wm^{-3}K^{-1}$ would fit better. While this is larger than the reported values in the literature, other effects may contribute to increasing *g*, such as interfacial contributions with the Au [40] or the oxide [41].

To further illustrate the effect of *g*, we plot in Figure 8 the evolution of *G* in the Al/sapphire system. In red we show the case of Al/Ta/sapphire, with a very rapid saturation within 1 nm. In blue and pink we artificially reduce the *g* of Ta by a factor of 100 and 1,350, respectively, to reach the *g* value of Au. We see that as *g* lowers, the saturation occurs at larger interlayer thickness beyond 12 nm. For comparison we also plot in black the Al/Au/sapphire system and show that the phonon dispersion of Au, with softer phonon modes than Ta, yields a slightly lower saturated *G* value, but the saturation sharpness is dominated by the *g* values.

### ***Presence of native oxide on Si substrates***

The films deposited on Si substrates include a native oxide layer, and it is therefore important to discuss here the role of this layer on the preceding results and analysis. Two aspects are treated here: the influence of the native oxide on the value for *G* measured and on the phonon transmission at the interface. While there is a relatively large variation in reported values for *G* at the Al/Si interface in the literature, recent evidence shows that the presence of the native oxide is compatible with the diffusive interface assumption of the DMM model and does not affect the interpretation of the results. For example, the values for *G* reported in samples with native oxide in the works of Duda *et al.* [42] and Gorham *et al.* [43] of ~190 and ~120 $MWm^{-2}K^{-1}$, respectively, are lower than what we report (245 $MWm^{-2}K^{-1}$). However, the data of Cheaito *et al.* [12] with native oxide gives a value of 220 $MWm^{-2}K^{-1}$, close to what we measure, and these are in reasonable agreement with the DMM predictions for the Al/Si interface. The work of Cheaito *et al.* for a broad range of interface metals on native oxide/Si substrates demonstrates that even with the presence of the oxide, important correlations for *G* can be made in agreement with DMM predictions.

Another aspect to consider is whether the presence of an oxide affects the phonon spectrum originating from the metal, as this can influence the phonon transmission or the diffusive vs. ballistic nature of the phonon transport. Hua *et al.* [44] show that the phonon transmission at the clean Al/Si interface appreciably differs from that having a native oxide only for phonons with wavelengths below ~1 nm. However, such phonon population does not primarily contribute to interface transport, as it is less populated and has lower group velocity. Additionally, Hua and Minnich [45] have shown that depending on the presence of a native oxide, the phonon flux at the Al/Si interface may have a different spectral contribution, which can lead to non-diffusive heat transport and breakdown of the interpretation of experimental data with a diffusive heat transport model. For example, samples without oxide lead to significant non-diffusive contributions to heat transport and pronounced optical spot size dependence of the derived Si thermal conductivity [46,47]. On the other hand, the presence of an oxide maintains the transport diffusive on a wide range of experimental modulation frequencies and optical spot sizes. In this work, the fitted Si thermal conductivity for all samples was 132 W/mK, close within experimental uncertainty to the expected bulk value, providing reassurance that non-diffusive phenomena are not playing a role. The same cannot be said for the work of Gorham *et al.* showing a lower value for $G$, as the authors point out.

***Influence of interfacial oxygen for ultrathin metallic layers***

The presence of oxygen in the native $SiO_2$ or $Al_2O_3$ substrate may lead to partial interfacial oxidation of some interlayer metals. One can gain insight by comparing the heat of formation of Ni, Ta, Al and Si oxides. The heat of formation of $SiO_2$ and $Al_2O_3$ (-859 kJ/mol and -1,676 kJ/mol) are much lower than NiO or $Ni_2O_3$ (-244 and 490 kJ/mol), indicating that oxygen would preferentially remain bonded in the $SiO_2$ layer or $Al_2O_3$ substrate rather than migrate to form Ni oxides in the Ni layer. This is confirmed in XPS measurements reported in the literature [48,49], and indicates that the thin Ni layers would indeed retain their metallic nature even at very low (<1 nm) film thickness. This implies that there should not be appreciable Ni oxide formation for the Ni interlayer deposition series. The heat of formation of $Ta_2O_3$, however, is -2,050 kJ/mol and is therefore lower than that of $SiO_2$ and $Al_2O_3$, indicating that some oxygen could migrate from the native $SiO_2$ or $Al_2O_3$ substrate to the Ta layer to form oxides. This is indeed confirmed by XPS [50], where metallic Ta is obtained when the film thickness is above 3 nm. The creation of a thin $TaO_x$ layer may provide an interpretation as to why the thickness dependence of $G$ in the Ta interlayer series saturates faster than predicted by our model. Nonetheless, the large *g* in Ta would guarantee that the value for *G* reaches saturation within 2 nm even in the metallic state.

**Conclusion**

In summary, we studied the thickness dependence of an interlayer at a metal-dielectric boundary. We demonstrate that very thin interface layers can alter the *G* at metal/dielectric interfaces considerably. The thickness dependent thermal boundary conductance in Al/Si and Al/sapphire changes significantly within the initial few nanometers and reaches saturation before ~5 nm on addition of Ni and Ta interlayers. When the interlayer has a strong electron-phonon coupling constant, the evolution of *G* is fast and happens within 2 nm. When *g* is weak, thermal boundary conductance saturates slowly. Thus, the electron-phonon coupling of the interlayer plays a major role in determining the trend of *G* evolution with thickness. The diffuse mismatch model can be adapted to model the thickness evolution of *G*, and an inelastic model that only considers acoustic phonon branches is found to be adequate in similarly matched interfaces, whereas an elastic acoustic model is better suited for mismatched interfaces.

**Acknowledgments**

The authors wish to acknowledge the Natural Sciences and Engineering Research Council of Canada, the Canada Foundation for Innovation, CMC Microsystems and York University for financial support.

**Supplementary Information**

The supplement presents the TDTR data and best fits for all the samples measured in this work, uncertainty analysis and error propagation.

Table 1: Thermal boundary conductance in the absence of the interlayers

| Interface | Al/Si | Al/sapphire |
|---|---|---|
| Acoustic + Optical modes<br>Inelastic scattering | 290 MW/(m²·K) | 410 MW/(m²·K) |
| Acoustic + Optical modes<br>Elastic scattering | 169 MW/(m²·K) | 238 MW/(m²·K) |
| Acoustic modes<br>Inelastic scattering | 245 MW/(m²·K) | 235 MW/(m²·K) |
| Acoustic modes<br>Elastic scattering | 169 MW/(m²·K) | 228 MW/(m²·K) |
| Experiment | 245 MW/(m²·K) | 205 MW/(m²·K) |

Table 2: Thermophysical properties of different materials of interest in this work. $\theta_D$ is the Debye temperature, $v_L$, $v_T$ and $v_{optical}$ are highest frequencies of the longitudinal optical, longitudinal transverse and optical phonon branches, respectively. $g$ is the electron-phonon coupling constant.

| Top Metal | Interlayer | Substrate |
|---|---|---|
| Al<br>$\theta_D$ = 428 K [51]<br>$v_L$= 9.6 THz [52]<br>$v_T$= 5.7 THz [52]<br>$g$ = 0.24×10$^{18}$ W/(m$^3$·K) [51] | Ni<br>$\theta_D$ = 450 K [51]<br>$v_L$ = 9.1 THz [29]<br>$v_T$ = 4.5 THz [29]<br>$g$ = 0.36 ×10$^{18}$ W/(m$^3$·K) [51] | α-$Al_2O_3$<br>$\theta_D$ =1035 K [53]<br>$v_L$ =10 THz [31]<br>$v_T$ = 6.9 THz [31]<br>$v_{optical}$ = 26 THz [31] |
| Au<br>$\theta_D$ = 165 K [51]<br>$v_L$ = 4.6 THz [54]<br>$v_T$ = 2.8 THz [54]<br>$g$ = 0.023×10$^{18}$ W/(m$^3$·K) [51] | Ta<br>$\theta_D$ = 225 K [55]<br>$v_L$ = 5.5 THz [28]<br>$v_T$ = 2.6 THz, 3.7 THz [28]<br>$g$ = 31×10$^{18}$ W/(m$^3$·K) [56] | Si<br>$\theta_D$ = 645 K [57]<br>$v_L$ = 12 THz [30]<br>$v_T$ = 4 THz [30]<br>$v_{optical}$ = 15.5 THz [30] |
|  | Cr<br>$\theta_D$ = 630 K [38]<br>$v_L$ = 10 THz [28]<br>$v_T$ = 6 THz, 7.7 THz [28]<br>$g$ = 0.42×10$^{18}$ W/(m$^3$·K) [58] |  |

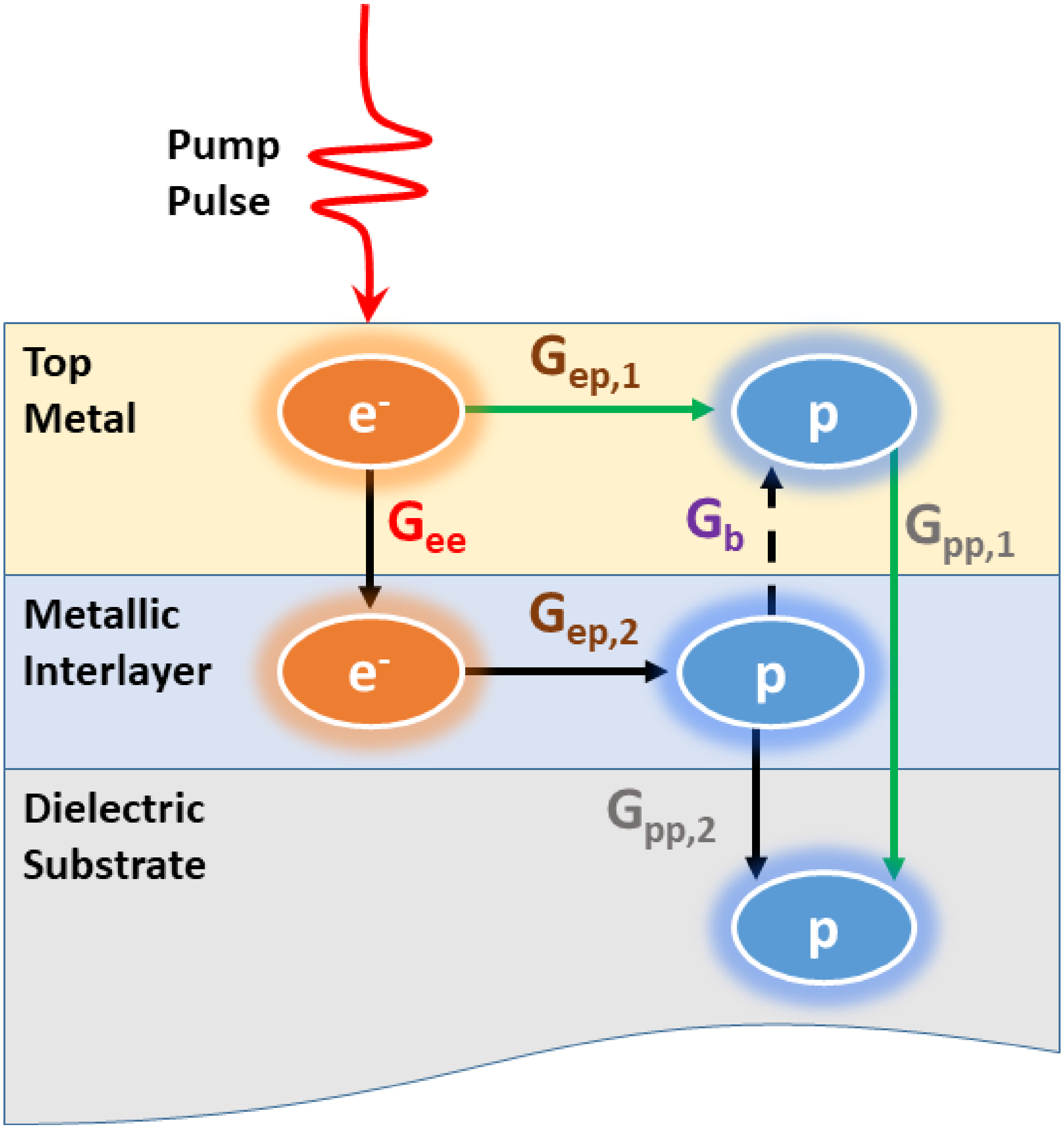

Figure 1: Thermal transport mechanisms at a metal-dielectric interface in the presence of an interlayer. Green and black solid arrows represent heat transport *pathway 1* and *pathway 2*, respectively. Black dashed arrow represents heat backflow from interlayer to metal.

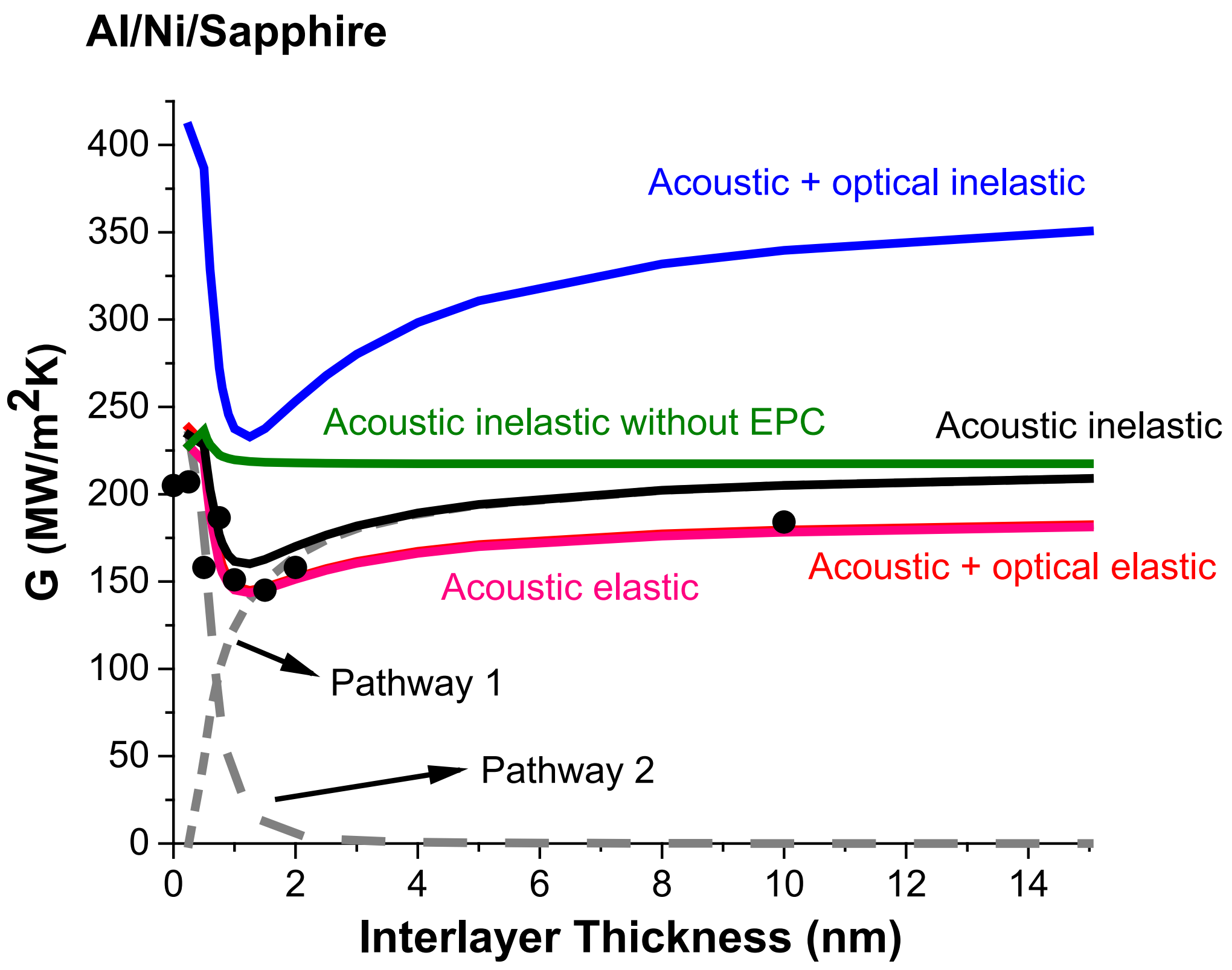

Figure 2: Comparison of experimental *G* with the model as a function of interlayer thickness for the Al/Ni/sapphire system. Experimental results are represented by filled circles. The black solid curve represents the total *G* evolution when considering all inelastic interactions with the acoustic phonons in sapphire. The dotted lines show the contributions due to *pathway 1* and *pathway 2* as a function of interlayer thickness. For comparison, the green curve shows the same model without considering the electron-phonon coupling *g* in the Ni layer. The red solid line denotes the model considering elastic interactions of all the phonon modes in sapphire. The blue solid curve shows the inelastic interactions of all phonon modes in sapphire. The pink curve is the model considering elastic processes with acoustic phonons in sapphire, and it is almost indistinguishable from the red curve.

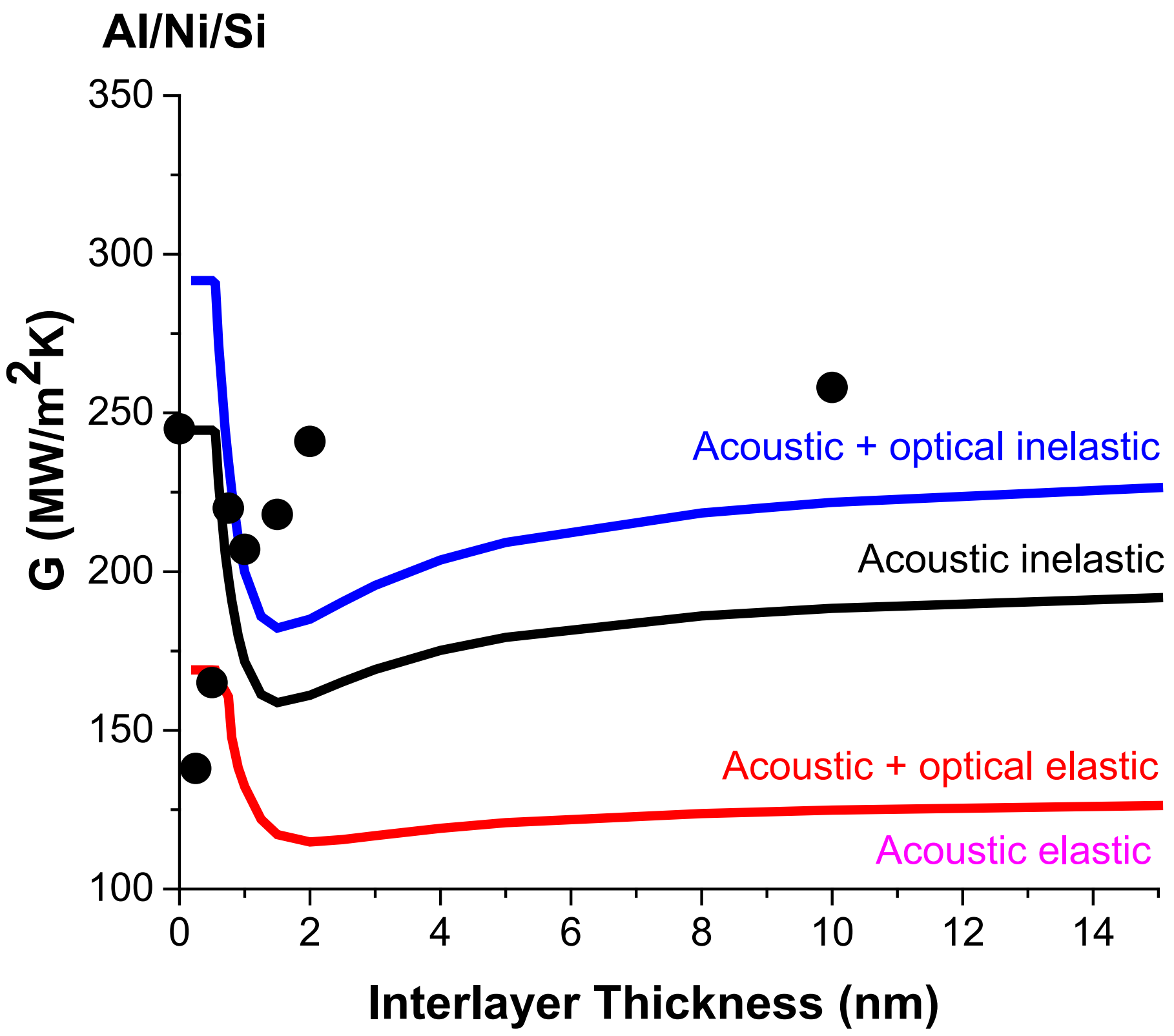

Figure 3: Comparison of experimental *G* with the model as a function of interlayer thickness for the Al/Ni/Si system. Experimental results are represented as filled circles. The black solid curve represents the total *G* evolution when considering all the inelastic interactions with the acoustic phonons in Si. The red solid line denotes the model considering elastic interactions of all the phonon modes in Si. The blue solid curve shows the inelastic interactions of all phonon modes in Si. Elastic processes with acoustic phonons in sapphire are almost indistinguishable from the red curve.

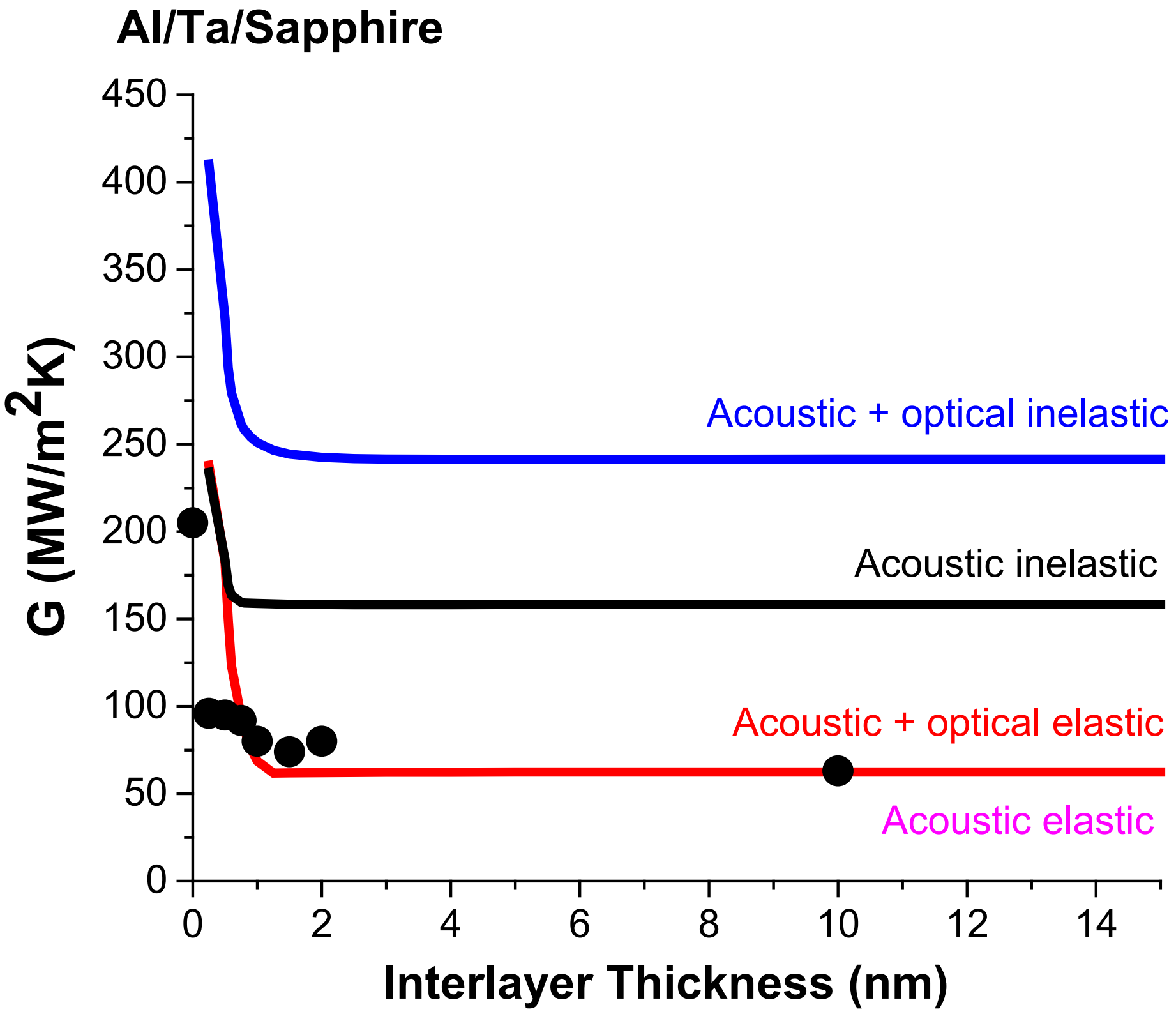

Figure 4: Comparison of experimental *G* with the model as a function of interlayer thickness for the Al/Ta/sapphire system. Experimental results are represented by filled circles. The black solid curve represents the total *G* evolution when considering all the elastic interactions with the acoustic phonons in sapphire. The red solid line denotes the model considering elastic interactions of all the phonon modes in sapphire. The blue solid curve shows the inelastic interactions of all phonon modes in sapphire. Elastic processes with acoustic phonons in sapphire *are* almost indistinguishable from the red curve.

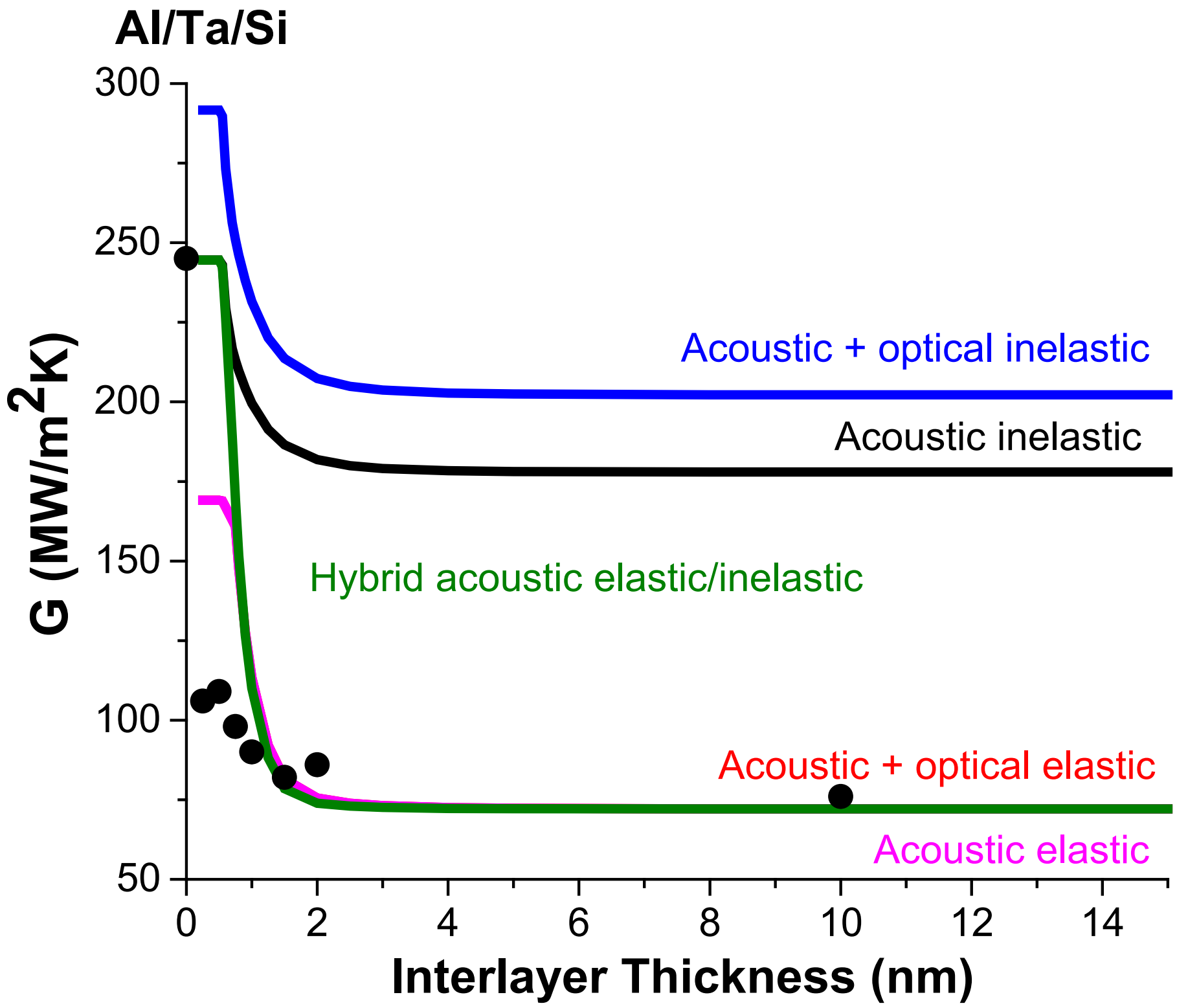

Figure 5: Comparison of experimental *G* with the model as a function of interlayer thickness for the Al/Ta/Si system. Experimental results are represented as filled circles. The black solid curve represents the total G evolution when considering all the elastic interactions with the acoustic phonons in Si. The red solid line denotes the model considering elastic interactions of all the phonon modes in Si. The blue solid curve shows the inelastic interactions of all phonon modes in Si. Elastic processes with acoustic phonons in sapphire are almost indistinguishable from the red curve. A hybrid elastic/inelastic model is shown in green.

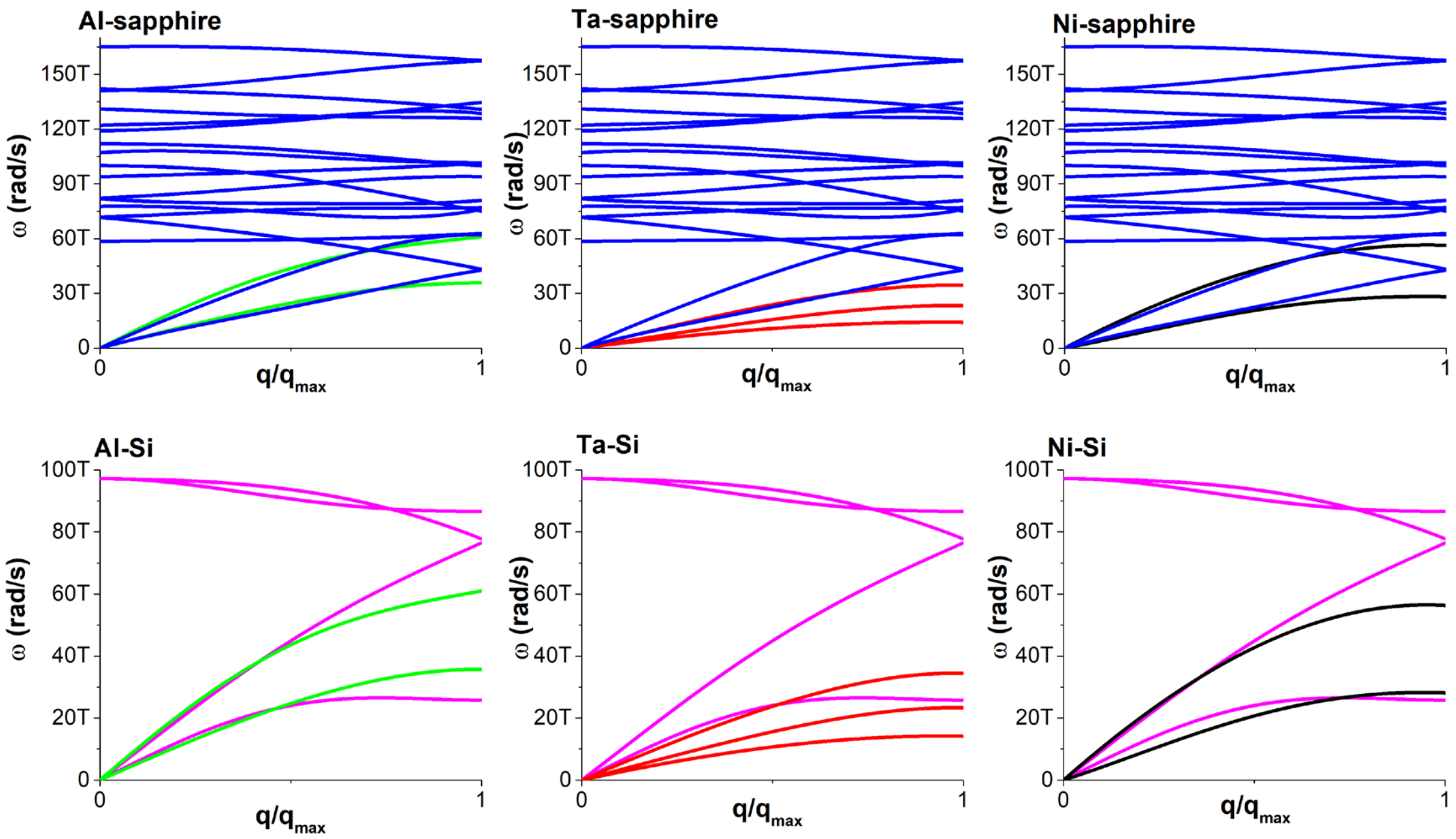

Figure 6: Phonon dispersions of metals and dielectric materials used in this work. Blue and pink curves represent the phonon dispersion branches for sapphire and Si, respectively. Green, red, and black represents the Al, Ta, and Ni phonon dispersion branches, respectively. Note that the *x*-axes of the dispersion curves are normalized for clarity. When estimating phonon energy overlaps, comparing the dispersion curves with a normalized *x*-axis is reasonable within a DMM model, since within the model's approximation momentum matching is not enforced in the calculation of the transmission coefficients.

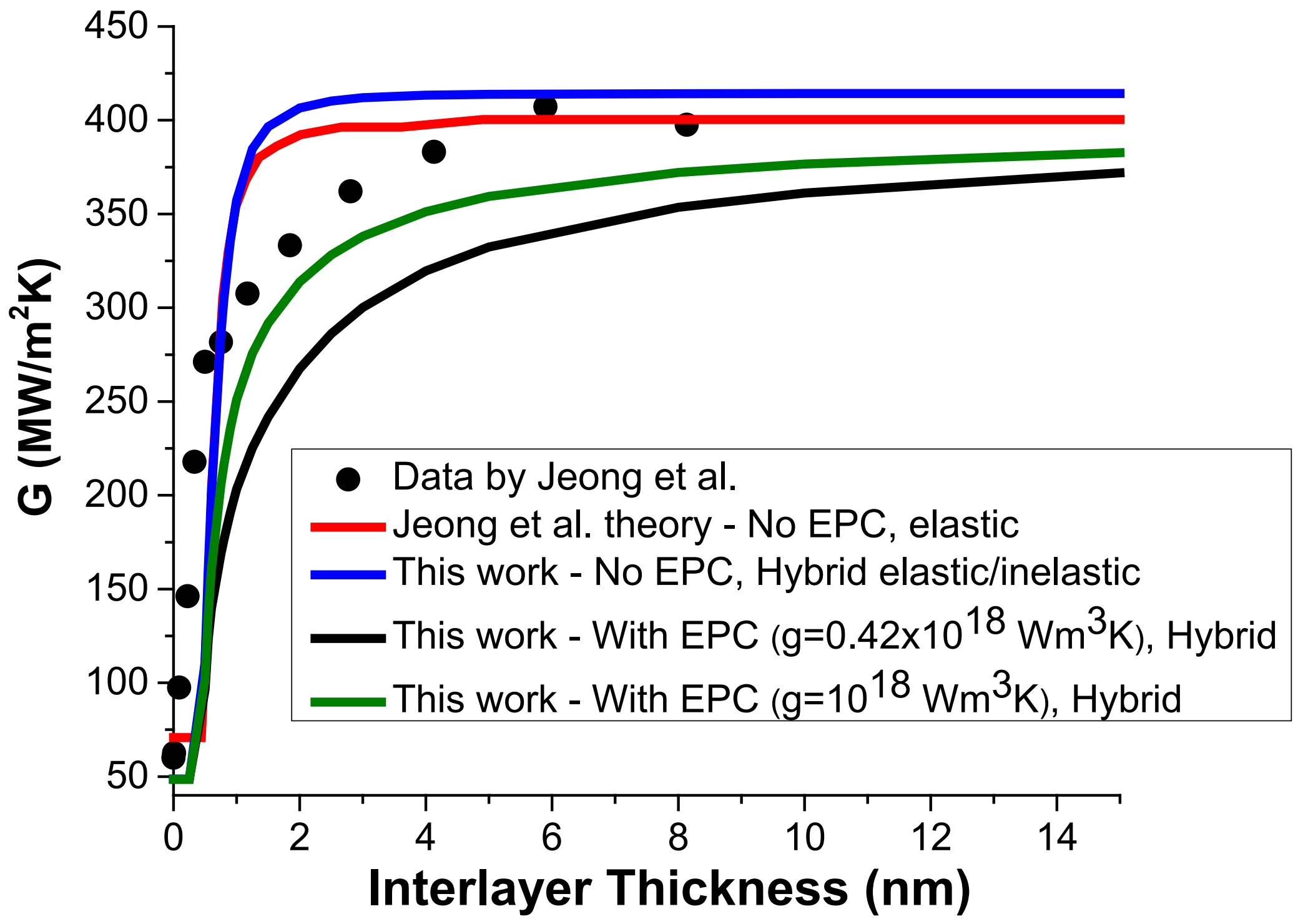

Figure 7: Comparison of our model with the results published in Ref. [15]

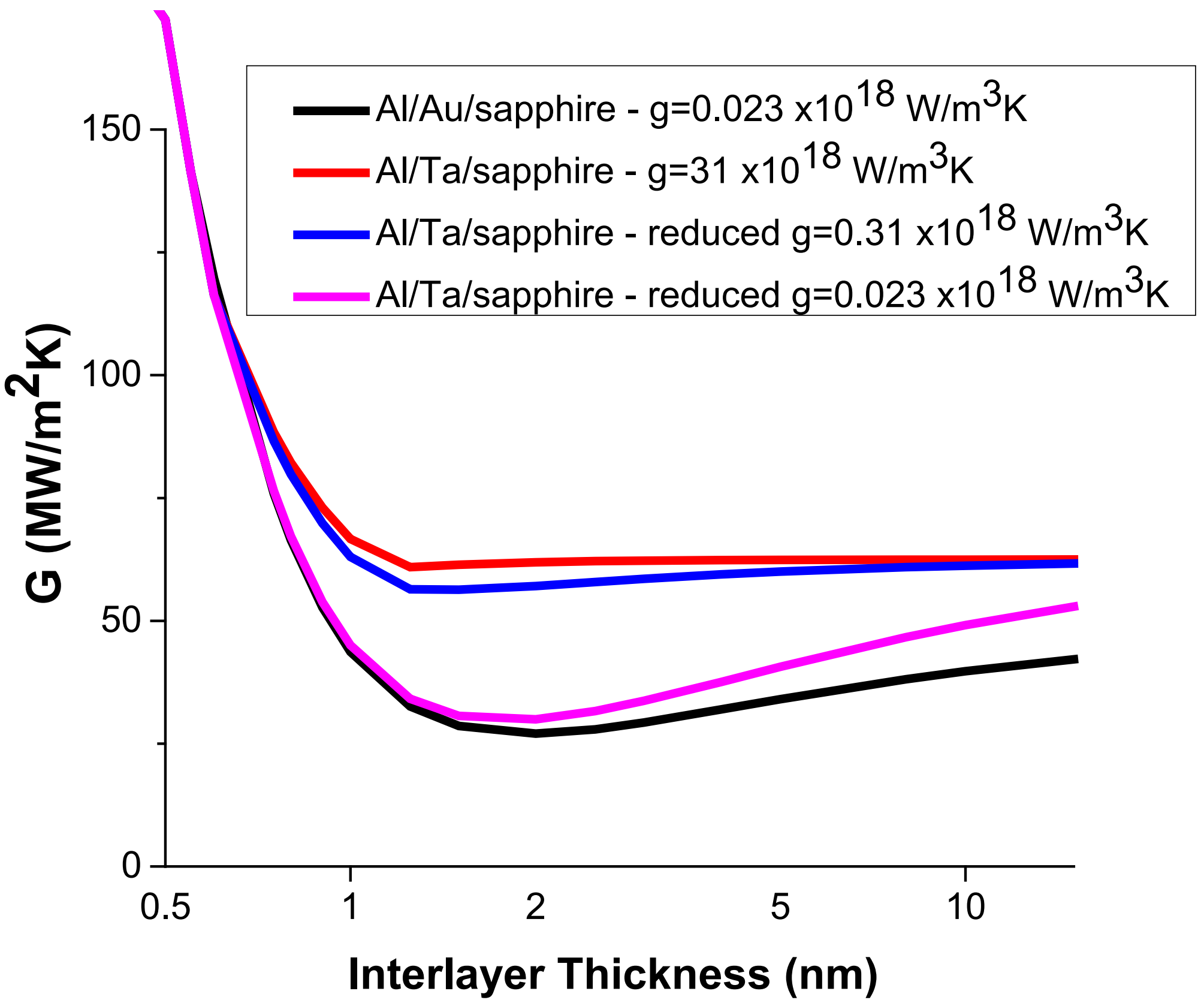

Figure 8: Comparison of modeled $G$ for the Al/Ta/sapphire system with different EPC values. The red, blue and pink curves show the effect of decreased $g$ values from the nominal Ta value of $31 \times 10^{18}$ W/($m^3$ ·K), and reduced to $0.31 \times 10^{18}$ W/($m^3$ ·K) and to the value of Au of $0.023 \times 10^{18}$ W/($m^3$ ·K). For comparison, the black curve is the Al/Au/sapphire system. The plot indicates that $g$ dominates how rapidly the TBC saturates.